\documentclass[aps,prl,twocolumn,superscriptaddress]{revtex4-1}
\usepackage{graphicx}
\usepackage{amsmath}
\usepackage{upgreek}
\usepackage{xcolor}
\newcommand{\ket}[1]{{\left| {#1} \right\rangle}}

\DeclareMathAlphabet{\mathbsf}{OT1}{cmss}{bx}{n}

\newcommand{\mrm}[1]{\mathrm{#1}}

\newcommand{\shrinkify}[1]{\textstyle {#1} \displaystyle}

\begin{document}
\title{Dipole-phonon quantum logic with trapped polar molecular ions}
\author{Wesley C. Campbell}
\author{Eric R. Hudson}
\affiliation{Department of Physics and Astronomy, Los Angeles, California 90095, USA}
\affiliation{UCLA Center for Quantum Science and Engineering, University of California – Los Angeles, Los Angeles, California 90095, USA}

\begin{abstract}
The interaction between the electric dipole moment of a trapped molecular ion and the configuration of the confined Coulomb crystal couples the orientation of the molecule to its motion.
We consider the practical feasibility of harnessing this interaction to initialize, process, and read out quantum information encoded in molecular ion qubits without optically illuminating the molecules.
We present two schemes wherein a molecular ion can be entangled with a co-trapped atomic ion qubit, providing, among other things, a means for molecular state preparation and measurement. 
We also show that virtual phonon exchange can significantly boost range of the intermolecular dipole-dipole interaction, allowing strong coupling between widely-separated molecular ion qubits.
\end{abstract}

\date{\today}
\maketitle

Trapped atomic ion systems have demonstrated the lowest state preparation and measurement (SPAM) infidelity~\cite{Christensen2019} and single- and two-qubit gate error rates~\cite{Harty2014,Balance2016,Gaebler2016} of any qubit. 
Fully programmable, few-qubit quantum computers based on trapped ions have already been constructed~\cite{Nigg2014, Debnath2016}. 
However, to date, these systems have not been scaled to a large number of qubits for reasons including anomalous heating~\cite{Brownnutt2015, Talukdar2016, Eltony2016, Sedlacek2018} and phonon mode crowding~\cite{Landsman2019}.

Recently, it was shown that molecular ion qubits, coupled by their direct, electromagneitc dipole-dipole interaction, could be used for quantum information processing~\cite{Hudson2018}. 
While the scalability of quantum molecular qubit systems is not expected to be limited by anomalous heating or phonon mode crowding, molecular ion qubits are not, currently, as easily controllable as atomic ion qubits.  
In particular, SPAM of molecular ions is made difficult by their typical lack of optical cycling transitions, which makes laser illumination of molecules problematic~\cite{Stollenwerk2018}. 
One approach is to perform quantum logic spectroscopy (QLS) via a co-trapped atomic ion \cite{Schmidt2005Spectroscopy,Wolf2016,Chou2017}. 
However, since QLS requires cooling near the ground state of motion, it is technologically demanding, and laser manipulation of molecular ions can lead to spontaneous emission to dark states.

\begin{figure}
\includegraphics[width=0.4\textwidth]{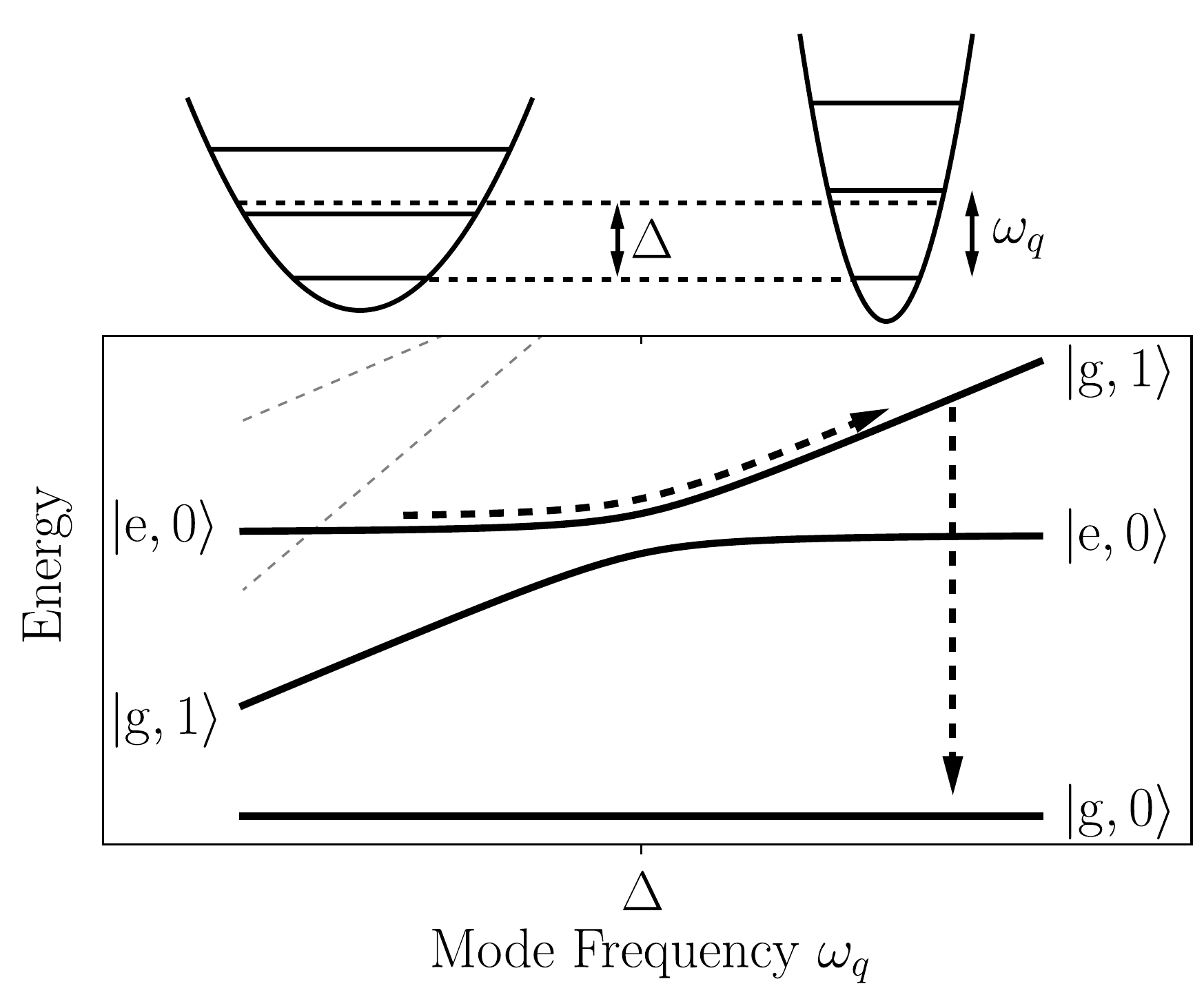}
\caption{Adiabatic dipole-phonon exchange.
Dressed energy levels of trapped molecular ion as a function of normal mode frequency $\omega_q$. State preparation can be accomplished by adiabatically increasing the trap frequency, which results in the conversion of $|\mrm{e},0\rangle$ to $|\mrm{g},1\rangle.$ The $|\mrm{g},1\rangle$ state can be detected by a sideband operation on a co-trapped atomic ion.}
\label{fig:arp}
\end{figure}

In this work, we describe how the dipole-phonon coupling in an ion trap can be used to entangle the dipole moment of a polar molecular ion with the phonon modes of a multi-ion Coulomb crystal.
This phenomenon can be intuitively understood in two ways: as the time-dependent electric field experienced by a non-stationary ion driving a molecular electric dipole transition, or as a time-dependent dipole moment driving ion motion.
For multiple ions, the oscillations occur in the collective modes of a Coulomb crystal, and even widely-separated dipoles can be made to interact strongly via shared phonon modes.
Further, the dipole-phonon interaction can be augmented by applying traditional sideband operations to a co-trapped atomic ion to effect molecule-atom entanglement and excitation exchange. This provides a potentially powerful means for non-destructive molecular ion SPAM that is similar to QLS, but can be free from the need to be in the motional ground state or to optically illuminate the molecules.
These operations form the basic building blocks of a hybrid atom-molecule quantum logic system, a means for coherent conversion between single microwave photons and atomic excitations, and a new route to quantum simulation of strongly-coupled matter.
In what follows, we first describe the basic dipole-phonon interaction and then treat several important cases of molecule-atom and molecule-molecule coupling. 

We consider a linear ion chain, which may contain both atomic and molecular ions of approximately the same mass (for simplicity, we assume they are equal), and treat motion only along the axial ($z$) direction of the chain.  
In a linear Paul trap, harmonic confinement in $z$ can be entirely electrostatic, allowing the use of a 1D static model.
Ion motion follows from the approximate Hamiltonian $\mathcal{H}_o = \sum_{p=1}^{N+M}\omega_p (a_p^\dagger a_p + \frac{1}{2})$, where $\omega_p$ is the frequency of normal mode $p$ and $N$ and $M$ are the number of molecular and atomic ions in the chain, respectively ($\hbar \equiv 1$). 
The displacement of ion $i$ from its equilibrium position can be written as a superposition of motion in the normal modes: $\hat{z}^{(i)} = \sum_{p=1}^N \mathrm{b}^{(i)}_p \hat{\zeta}_p$.  Here, $\hat{\zeta}_p$ is the displacement operator of the normal mode $p$ and we take the normal mode eigenvectors $\mathbf{b}_p$ to be normalized and real.
Thus, $\hat{z}^{(i)} = \sum_{p=1}^N  \sqrt{\frac{1}{2 m \omega_p}} \mathrm{b}_p^{(i)}(a_p + a_p^\dagger)$~\cite{James1998}.

Molecular ions in the chain are assumed to be identical polar molecules, each with a pair of opposite-parity states $|\mathrm{g}^{(i)}\rangle$ and $|\mathrm{e}^{(i)} \rangle$ that represent the -1 and +1 eigenstates, respectively, of the Pauli operator $\sigma_Z^{(i)}$ for this effective two-level system of molecule $i$.  
The molecular states are separated in energy by the noninteracting Hamiltonian $\mathcal{H}_\mathrm{m}^{(i)} = \frac{\Delta}{2}\sigma_Z^{(i)}$ and are connected by an electric dipole transition moment $d = |\langle \mathrm{g} |\mathbf{d}\cdot \hat{\mathbf{z}}| \mathrm{e}\rangle |$.
Physically, these states could be any dipole-connected states, such as rotational states or a $\Lambda$-, $\Omega$-, or $l$-doublet, and they define the Hilbert space of the dipole.

If the transition frequency between dipole states is close to any of the normal mode frequencies, interactions between the dipole and phonons become important. 
This interaction follows from $\mathcal{H}_\mrm{dp}^{(i)} = -\mathbf{d}^{(i)}\cdot\mathbf{E}^{(i)}$, where $\mathbf{d}^{(i)} = d \,\sigma_X^{(i)} \hat{\mathbf{z}}$ is the dipole moment and $\mathbf{E}^{(i)} = E(\hat{z}^{(i)})\hat{\mathbf{z}}$ is the instantaneous total electric field (both the static trap field and the Coulomb repulsion of all other ions in the trap) at the position of ion $i$. 
The total electric field experienced by ion $i$ can be found from the effective potential  $V = \sum_{p=1}^N V(\hat{\zeta}_p)$~\cite{James1998} as $E^{(i)} = -\frac{1}{e}\sum_{p = 1}^N\frac{\partial V(\hat{\zeta}_p)}{\partial \hat{\zeta}_p}\frac{\partial \hat{\zeta}_p}{\partial \hat{z}^{(i)}} =  -\frac{m}{e}\sum_{p = 1}^N \omega_p^2\mathrm{b}^{(i)}_p \hat{\zeta}_p$.
The characteristic length scale for each mode, $\ell_p \equiv \left(e^2/(2 \pi \epsilon_\mathrm{o} m \omega_p^2)\right)^\frac{1}{3}$, which is approximately the ion spacing, can be used to rewrite the electric field operator in the familiar form
\begin{equation}
    E^{(i)} = -\sum_{p=1}^N  \sqrt{\frac{\omega_p}{2 \epsilon_\mathrm{o} V_p}}(a_p + a^\dagger_p)\mathrm{b}_p^{(i)} \label{eq:QuantumOpticsE}
\end{equation}
where the effective mode volume is $V_p \equiv 2 \pi \ell_p^3$.  Equation (\ref{eq:QuantumOpticsE}) closely resembles the quantized electromagnetic field used in quantum optics, and we can treat phonons as if they were photons, in the dipole aproximation, confined to mode volume $V_p$ with spatial dependence given by $\mathrm{b}_p^{(i)}$.

Using this, the dipole phonon interaction is $\mathcal{H}_t^{(i)} = \frac{d}{e} \sqrt{\frac{m}{2}}\sigma_X^{(i)} \sum_{p=1}^N  \omega_p^{3/2} \mathrm{b}_p^{(i)}(a_p + a_p^\dagger)$.
Identifying the vacuum Rabi frequency as $g_p^{(i)} \equiv d\, \mathcal{E}_{\mathrm{o},p}^{(i)} = d \sqrt{\frac{2 \omega_p}{ \epsilon_\mathrm{o} V_p}}\,\mathrm{b}_p^{(i)}$, where $\mathcal{E}_{\mathrm{o},p}^{(i)}$ is the electric field amplitude at ion $i$ from a single phonon in mode $p$,  the dipole-phonon interaction can be written as 
\begin{equation}
\mathcal{H}_\mrm{dp}^{(i)} = \sum_{p=1}^N\frac{g_p^{(i)}}{2}(a_p + a_p^\dagger) \sigma_X^{(i)}. \label{eq:DipPhonHam}
\end{equation}

To produce coupling between molecules and atoms, an atomic qubit ion in the linear chain may be addressed by laser fields to couple it to motion.
For atomic ion $j$, this interaction is described by the Hamiltonian $\mathcal{H}_\mrm{a}^{(j)} = \frac{\omega_\mathrm{a}}{2}s_Z^{(j)} + \frac{\Omega^{(j)}}{2} \left(s_+^{(j)}e^{-\imath\omega_L t}  + s_-^{(j)}e^{\imath\omega_L t} \right)\left(1 + \sum_p \mathrm{b}_p^{(j)}\eta_p\left(a_p + a^\dagger_p\right)\right),$ where $\omega_\mathrm{a}$ is the energy splitting of the atomic qubit, $s_R^{(j)}$ are the Pauli operators on the atomic qubit, $\Omega^{(j)}$ is the (carrier) Rabi frequency of the laser-qubit interaction, $\omega_L$ is the laser (or beatnote) frequency, and $\eta_p$ is the Lamb-Dicke factor for mode $p$. 
Thus, the total Hamiltonian for a linear chain composed of $N$ identical polar molecules and $M$ identical atoms is 
\begin{equation}
\mathcal{H} =  \mathcal{H}_{\mrm{o}} + \sum_i^N(\mathcal{H}_\mrm{m}^{(i)} + \mathcal{H}_\mrm{dp}^{(i)}) + \sum_j^M\mathcal{H}_\mrm{a}^{(j)}.
\label{eqn:totalHamiltonian}
\end{equation}

In what follows, we consider several realizations of a chain of molecular and atomic ions and analyze the behavior that follows from Eq.~\ref{eqn:totalHamiltonian}. Unless specified otherwise, all numerical examples assume a molecular species (based loosely on $\mathrm{DCl}^+$ \cite{Brown1979Effective}) with the following properties: $m=37\mbox{ amu}$, $d=1\mbox{ D}$, $\Delta/2 \pi = 5\mbox{ MHz}$ in a trap with axial, center of mass frequency $\omega_1/2 \pi = 5 \mbox{ MHz}$.

\textit{Near-resonant dipole-phonon exchange} -- If the energy separation of the dipole states $\Delta$ is near that of the $q$th normal mode, and far from all other modes $p\neq q$, the rotating wave approximation can be applied to the molecular terms in Eq.~\ref{eqn:totalHamiltonian} to yield the Hamiltonian for molecule $i$ and phonon mode $q$,
\begin{equation}
\mathcal{H}^{(i)}_q \approx \mathcal{H}_\mathrm{m}^{(i)} +  \omega_q\left( a_q^\dagger a_q + \shrinkify{\frac{1}{2}} \right)   + \frac{g_q^{(i)}}{2} \left(a_q \sigma_+^{(i)}  + a_q^\dagger \sigma_-^{(i)} \right).
\label{eqn:hamiltonian}
\end{equation}
Except for the dependence of $g_q^{(i)}$ on $\omega_q$, this is the well-studied Jaynes-Cummings Hamiltonian~\cite{Jaynes1963} and thus immediately suggests several methods for dipole-phonon quantum logic.
For example, if $\omega_q=\Delta$, a molecule prepared in the state $\ket{\mathrm{e}_\mathrm{m},0}$ (${}_\mathrm{m}$ and ${}_\mathrm{a}$ denote molecule and atom, respectively) will undergo
vacuum Rabi flopping to $\ket{\mathrm{g}_\mathrm{m},1}$ at the rate $g_q$, and both excitation exchange as well as entanglement between the dipole and the phonon mode can be created by choosing the time for which the coupling is active ($\omega_q=\Delta$).

This interaction provides a robust means for SPAM of the molecular ion.
The coupling $g_q^{(i)}$ causes an an avoided crossing between states $|\mathrm{g}_\mathrm{m}^{(i)}\!,n_q\!+\!1 \rangle$ and $|\mathrm{e}_\mathrm{m}^{(i)}\!,n_q\rangle$ at $\omega_q = \Delta$, as shown in Fig.~\ref{fig:arp}(c).
By using a co-trapped atomic ion to laser cool to the ground state of motion and then adiabatically performing a linear sweep of the trap frequency $\omega_q$ through $\Delta$ from below, a polar molecule in $|\mathrm{e}_\mathrm{m}^{(i)},0 \rangle$ will emit a phonon in the shared motional mode $q$ with probability $P\approx 1-\exp (-2\pi g_q^{(i)2}/\dot{\omega}_q )$ and be transferred to $|\mathrm{g}_\mathrm{m}^{(i)}\!,1 \rangle$.  
Subsequent measurement of the phonon state followed by ground state cooling via the atomic ion would then prepare $|\mathrm{g}_\mathrm{m}^{(i)}\!,0 \rangle$ in a manner similar to traditional QLS~\cite{Wolf2016,Chou2017}. 
Repeating this scheme following a $\pi$-pulse on the $\ket{\mathrm{e}_\mathrm{m}}\leftrightarrow\ket{\mathrm{g}_\mathrm{m}}$ transition allows discrimination of $\ket{\mathrm{e}_\mathrm{m}}$ and $\ket{\mathrm{g}_\mathrm{m}}$ as a molecule initially in $\ket{\mathrm{e}_\mathrm{m}}$ emits a phonon on both ramps, a molecule initially in $\ket{\mathrm{g}_\mathrm{m}}$ emits a phonon only on the second ramp, and a molecule in a state outside of the $\ket{\mathrm{e}_\mathrm{m}}$-$\ket{\mathrm{g}_\mathrm{m}}$ subspace does not emit a phonon on either ramp.
Further, using circularly polarized radiation for the $\pi$-pulse in combination with multiple ramp sequences could `acoustically pump' the polar molecule into a single Zeeman sublevel.

This technique appears to have several advantages over traditional QLS. 
For example, here the dipole-phonon exchange does not require optical addressing of the molecule (which can lead to spontaneous emission) as $\omega_q$ can be tuned by simply changing DC trap voltages. 
Further, the dipole-phonon interaction has no carrier transition, which allows this sweep to span a wide range with strong coupling.

For state preparation, it is also possible to take a dissipation engineering approach, where ground-state (sideband or EIT) cooling is carried out while $\omega_q = \Delta$. 
At this operating condition, the dipole-phonon interaction couples the internal states of the molecule to the atomic spontaneous emission providing controllable dissipation in the $\ket{\mathrm{e}_\mathrm{m}}$-$\ket{\mathrm{g}_\mathrm{m}}$ subspace.
Thus, the $\ket{\mathrm{g}_\mathrm{m},0}$ state is prepared simply by cooling the atom while $\omega_q = \Delta$, and success can be verified using the adiabatic ramp sequence defined above for state detection.

\textit{Resonant Dipole-phonon exchange with multiple molecular ions} -- For multiple molecular ions in the trap, the Hamiltonian is $\mathcal{H} = \sum_i \mathcal{H}_q^{(i)}$, (see Eq.~(\ref{eqn:hamiltonian})).
Because the molecular ions couple to a shared phonon mode, it is straightforward to entangle them.
Using the center-of-mass mode so that all ions possess the same $\mathrm{b}_1^{(i)}$, a system initially prepared in the state $|\mathrm{g},\mathrm{g},...,\mathrm{g},1 \rangle$ will evolve as $\ket{\psi(t)} = \cos(\frac{g_q t}{\sqrt{N}}) |\mathrm{g},\mathrm{g},...,\mathrm{g},1 \rangle + \imath \sin (\frac{g_q t}{\sqrt{N}}) \ket{\mathrm{W},0}$, where 
$\ket{\mathrm{W}} = \frac{1}{\sqrt{N}}(\ket{\mathrm{e},\mathrm{g},\mathrm{g},...,\mathrm{g}} + \ket{\mathrm{g},\mathrm{e},\mathrm{g},...,\mathrm{g}} + ... + \ket{\mathrm{g},\mathrm{g},\mathrm{g},...,\mathrm{e}})$. 
Therefore, a molecular W state can be produced by either tuning the phonon mode on resonance with the molecule for a time $t_G = \sqrt{N}\pi/(2g_q)$ or by adiabatic rapid passage through the $\omega_1 = \Delta$ condition. 

This basic scheme provides a number of useful capabilities.
As a numerical example,
for two molecular ions, the gate time is $t_G \approx 25~\upmu$s to produce the Bell state $\ket{\psi^+} = \frac{1}{\sqrt{2}}(\ket{\mathrm{g},\mathrm{e}}+\ket{\mathrm{e},\mathrm{g}})$. 
This is at least two orders of magnitude faster than gate time for producing the same entangled state via the direct electromagnetic dipole-dipole coupling~\cite{Hudson2018}.

This interaction also provide a means for fast, high-fidelity quantum transduction from a microwave photon qubit into an atomic ion qubit or optical photon.
By placing an ion chain (containing many molecules and one atomic ion) in the mode of a microwave resonator whose frequency is at $\Delta$~\cite{Schuster2011}, the strong coupling limit of cavity QED can be reached allowing a microwave photon to be coupled to the collective dipole, creating $\ket{\mathrm{W}}$. 
Using the interaction described here, the $\ket{\mathrm{W}}$ state can then be transferred to motion, and the motion transferred to the state of the atom, which can be converted into optical-frequency photons, if desired.

\textit{Long-range dipole-dipole interactions mediated by virtual phonons} -- When all normal mode frequencies are detuned from the molecular dipole splitting $\Delta$, off-resonant absorption and emission of phonons still mediates effective dipole-dipole interactions between co-trapped molecules.  
Virtual phonon exchange occurs through every normal mode, and the effective Hamiltonian \cite{James2007effective} takes the form
\begin{eqnarray}
    \mathcal{H}_\mathrm{eff} &=& \sum_{i>j} J_{ij} \left(\sigma_+^{(i)}\sigma_-^{(j)} + \sigma_-^{(i)}\sigma_+^{(j)} \right) \label{eq:Term1}\\
    &  &+ \sum_{i,p}\frac{2\Delta}{\Delta^2 - \omega_p^2}\left(\frac{g_p^{(i)}}{2}\right)^2 \left(a_p^\dagger a_p + \shrinkify{\frac{1}{2}} \right)\sigma_Z^{(i)}\label{eq:Term2}
\end{eqnarray}
where the effective dipole-dipole interaction strength is given by
\begin{eqnarray}
    J_{ij}& =& \sum_p \frac{2 \omega_p}{\Delta^2 - \omega_p^2}\left(\frac{g_p^{(i)}g_p^{(j)}}{4}\right)\label{eq:Jij} \\
    & = & \sum_p \left(\frac{d^2}{2 \pi \epsilon_\mathrm{o} \ell_p^3} \right)\frac{\omega_p^2}{\Delta^2 - \omega_p^2} ~\mathrm{b}_p^{(i)}\mathrm{b}_p^{(j)}.\label{eq:JijDeconstructed}
\end{eqnarray}

\begin{figure}
\includegraphics[width=0.45\textwidth]{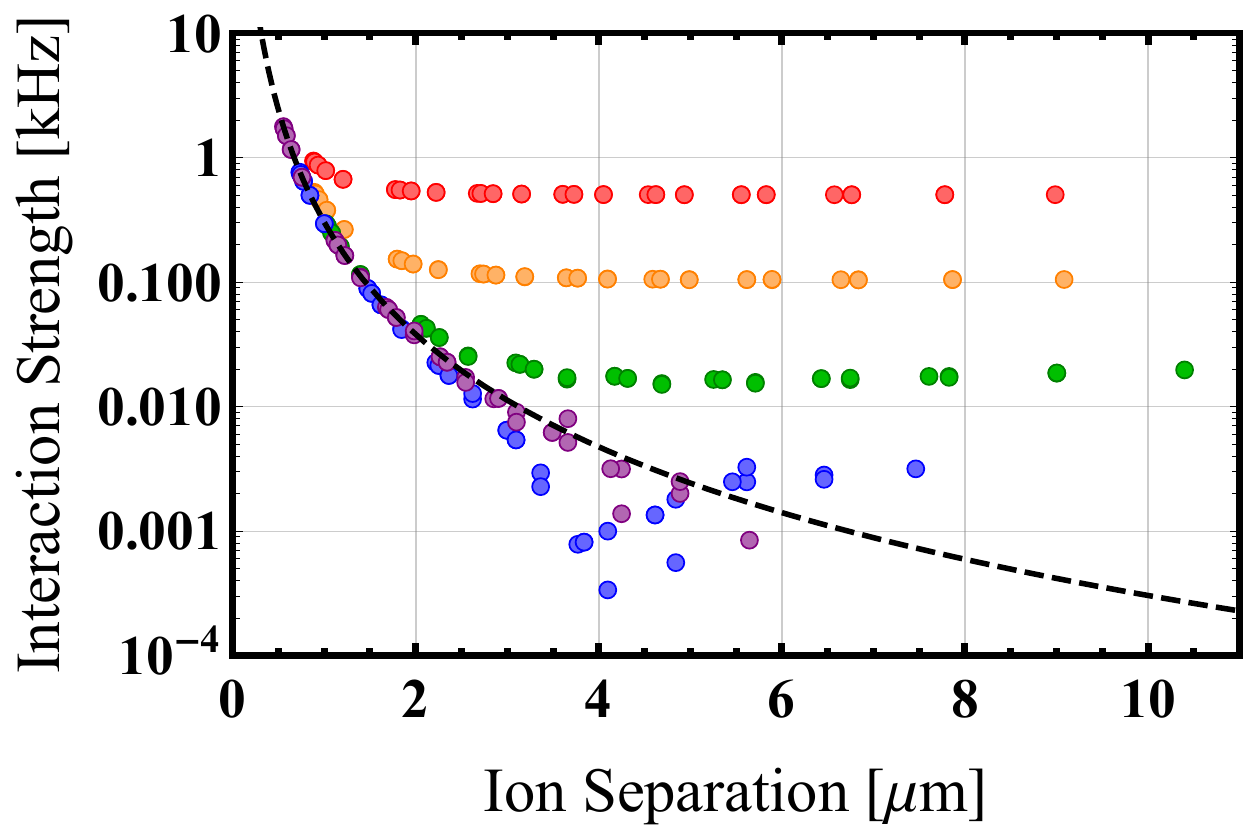}
\caption{Calculated strength of the phonon-mediated dipole-dipole interaction (from Eq.~\ref{eq:Jij}) as a function of the distance between 10 molecular ions in a chain.  The red curve shows the direct, electromagnetic dipole-dipole interaction strength.  Circles show the calculation for $\omega_1/2\pi$ = ($4.98$, $4.90$, $4.00$, $6.58$, $10.00$) MHz for the (red, orange, green, blue, purple) circles.  The black, dashed curve is the direct, electromagnetic dipole-dipole interaction,  illustrating that the phonon-mediated dipole-dipole interaction can be made longer range and orders of magnitude stronger.}
\label{fig:Range}
\end{figure}

\begin{figure}
\includegraphics[width=0.4\textwidth]{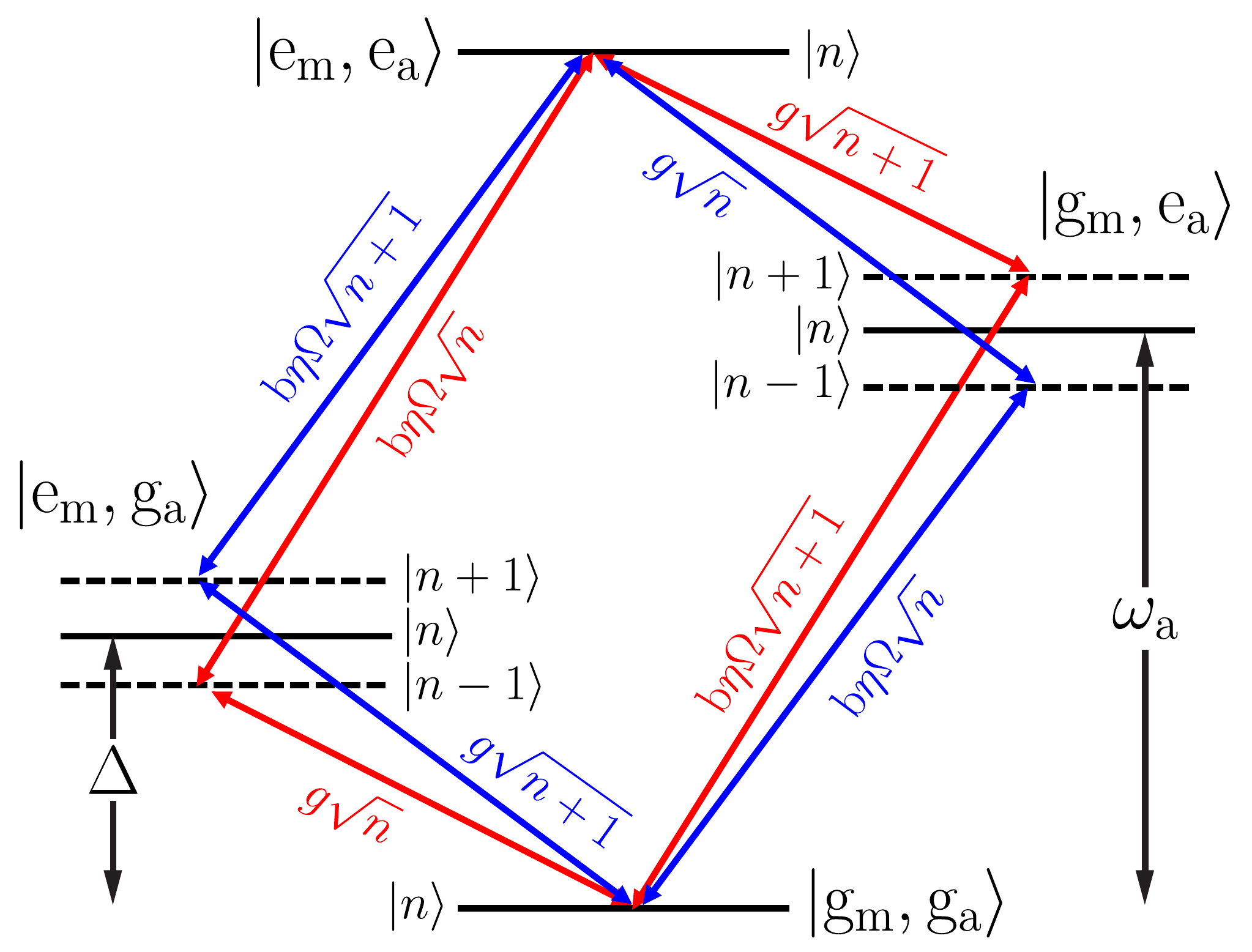}
\caption{Energy levels and couplings of a co-trapped atomic and molecular ion while driving the atomic ion with a laser at $\omega_L = \omega_\mathrm{a}' + \Delta$. Multiple pathways interfere to remove the $n$-dependence of the coupling between $\ket{\mathrm{g}_\mathrm{m},\mathrm{g}_\mathrm{a},n}$ and $\ket{\mathrm{e}_\mathrm{m},\mathrm{e}_\mathrm{a},n}$, allowing quantum control without the need for cooling to the Lamb-Dicke regime.}
\label{fig:nIndepFig}
\end{figure}

The dipole-dipole term (\ref{eq:Term1}) is an exchange (or XY) interaction, similar to the electromagnetic dipole-dipole interaction \cite{Hudson2018}.  A similar effective Hamiltonian has been shown to arise when atomic ions are driven with a single sideband \cite{Senko2015Realization}, and the validity of that approximation has been shown to hold in various parameter regimes \cite{Wall2017Bosonmediated}.
The functional form of the coupling strength, Eq.~(\ref{eq:Jij}) (see Fig.~\ref{fig:Range}), is similar to the phonon mode dependence of the Ising interaction that arises for atomic ions driven by a bichromatic force \cite{MolmerSorensen99Multiparticle,Porras2004Effective,James2007effective}, but with a stronger dependence on $\omega_p$.  
Likewise, and for similar reasons (see Fig.~\ref{fig:nIndepFig}), it is insensitive to temperature, and should work outside of the Lamb-Dicke regime. 
By writing it in the form (\ref{eq:JijDeconstructed}), it can be seen that the strength of the phonon-mediated dipole-dipole interaction (for phonon mode frequency $\omega$) can be stronger than the direct, electromagnetic interaction by a factor of approximately $\mathrm{b}^{(i)}\mathrm{b}^{(j)}\omega^2/(\Delta^2 - \omega^2)$.

Figure~\ref{fig:Range} shows calculated values of $|J_{ij}|$ as a function of the separation between molecular ions in a 10-ion, harmonically confined linear chain.  
As the center-of-mass trap frequency $\omega_1$ is moved further from $\Delta$ (red$\rightarrow$purple), the short range behavior approaches the direct, electromagnetic dipole-dipole interaction (black, dashed curve, $d^2/2 \pi \epsilon_\mathrm{o}r^3$).  In all cases shown in Fig.~\ref{fig:Range}, the phonon mediation produces dipole-dipole interactions that exceed their electromagnetic counterpart by orders of magnitude at long range, providing a potential new path to the study of strongly-coupled quantum systems. 
 
The second term in the effective Hamiltonian (\ref{eq:Term2}) is a ``quantum AC Stark shift'' \cite{James2007effective,DHelon1996Measurements} of the phonon and dipole frequencies.
While the effects of this term may be mitigated by a spin echo sequence \cite{Wall2017Bosonmediated}, it provides another means for state detection since it allows the state of the molecule to be determined by simply measuring a shift in the normal mode frequencies.  For $\Delta/2 \pi$ = 4.98\mbox{ MHz}, the center of mass mode frequency shift is $\approx\! 2.5\mbox{ kHz}$.

\textit{Long-range dipole- atomic qubit interaction mediated by virtual phonons} -- An atomic qubit, driven by an appropriate laser frequency, can be made to participate in the interaction just described.

If the atomic qubit laser is applied at $\omega_L=\omega_\mathrm{a}^\prime+\Delta$ (where $\omega_\mathrm{a}^\prime$ is the atomic level splitting $\omega_\mathrm{a}$ with the AC Stark shift of the laser included), it is blue detuned with respect to the atomic qubit and (off-resonant) excitation of the atom is most likely to be accompanied by the creation of a phonon, shown as the right-hand, red path in Fig.~\ref{fig:nIndepFig}.  
This phonon can then be absorbed by the molecule, and 
the transition amplitude for this path is proportional to $n+1$.  
Likewise, this process can occur in the reverse order (left red path in Fig.~\ref{fig:nIndepFig}) with transition amplitude $\propto -n$, and the $n$-dependence of the overall amplitude for the $\ket{\mathrm{g}_\mathrm{m},\mathrm{g}_\mathrm{a},n} \leftrightarrow \ket{\mathrm{e}_\mathrm{m},\mathrm{e}_\mathrm{a},n}$ transition drops out.
The blue lines in Fig.~\ref{fig:nIndepFig} indicate terms that are further off resonance, for which the motional state dependence also cancels.  
As a result, the effective Rabi frequency for oscillations from $\ket{\mathrm{g}_\mathrm{m},\mathrm{g}_\mathrm{a},n}$ to $\ket{\mathrm{e}_\mathrm{m},\mathrm{e}_\mathrm{a},n}$ is $\Omega_{\rm{eff}} \approx \mathrm{b}_q\eta_q \Omega g_q/2(\Delta-\omega_q)$ and is independent of the motional state of the system. 
The case for $\omega_L=\omega_\mathrm{a}^\prime-\Delta$ (i.e. laser red detuned with respect to the qubit) is the same with the substitution $\mathrm{g}_\mathrm{a} \leftrightarrow \mathrm{e}_\mathrm{a}$, and can therefore be used to drive transitions $\ket{\mathrm{g}_\mathrm{m},\mathrm{e}_\mathrm{a},n} \leftrightarrow \ket{\mathrm{e}_\mathrm{m},\mathrm{g}_\mathrm{a},n}$ in the same manner.  

Along with single ion rotations, this toolbox allows full control of the Hilbert space of the dipole orientation and atomic qubit, and can be used to create entangled states of the hybrid molecule atom system without the need to operate in the Lamb-Dicke limit.
For example, this interaction can be used to rotate the system into an atom-molecule GHZ state, $\frac{1}{\sqrt{2}}(|\mrm{g}_\mrm{m},\mrm{g}_\mrm{a}\rangle + |\mrm{e}_\mrm{m},\mrm{e}_\mrm{a}\rangle)$, in a time $t_\mathrm{G} = \frac{2\pi(\Delta-\omega_q)}{\mathrm{b}_q\eta_q\Omega g_q}$, which for the example molecule described above and $\eta_q \Omega/2 \pi = 100\mbox{ kHz}$ is $t_\mathrm{G}=20\mbox{ }\upmu\mbox{s}$.

\textit{Molecules amenable to dipole-phonon quantum logic}-- For these ideas to be practically feasible with current ion trap technology requires that $\Delta/2 \pi \leq 20$~MHz.
For diatomic molecules, the most likely targets are $\Lambda$-$/\Omega$-doublet ground states, which have opposite parity and can be closely spaced, such as the $\mrm{X}{}^2\Pi_{3/2}$ state of $\mathrm{DCl}^+$, which has $\Delta/2 \pi = 8.3\mbox{ MHz}$ \cite{Brown1979Effective}. 
Other examples include HfF$^+$ and ThF$^+$, which are relevant for electron electric dipole moment searches and have $\Delta/2 \pi\approx~740$~kHz and $5.3$~MHz~\cite{Cossel2012,Gresh2016}.
For linear polyatomic molecules,  many $\ell$-doublets are split by $\lesssim10$~MHz~\cite{Nielsen1943,Kozyryev2017,Puri2017}.
These are especially interesting candidates as they may also possess diagonal Franck-Condon factors~\cite{Kozyryev2017, Orourke2019}, allowing, among other things, state preparation via optical pumping~\cite{Lien2014}.

In summary, we have described the interaction between the dipolar degree of freedom of a molecular ion with the phonon modes in an ion trap and its use for quantum logic. Specifically, we have shown that this interaction can be used for state preparation and measurement of molecular ions, entanglement of molecular ions, entanglement of molecular ions with atomic ions, the transduction of quantum information from the microwave to optical regime, and the study of strongly correlated quantum matter.  In many cases, these applications do not require cooling to the ground state of motion.

\begin{acknowledgments}
The authors acknowledge David Patterson for helpful discussions. This work was supported by the U.S. Department of Energy, Office of Science, Basic Energy Sciences, under Award \#DE-SC0019245.
\end{acknowledgments}

\bibliography{DipolePhonon}

\begin{thebibliography}{33}%
\makeatletter
\providecommand \@ifxundefined [1]{%
 \@ifx{#1\undefined}
}%
\providecommand \@ifnum [1]{%
 \ifnum #1\expandafter \@firstoftwo
 \else \expandafter \@secondoftwo
 \fi
}%
\providecommand \@ifx [1]{%
 \ifx #1\expandafter \@firstoftwo
 \else \expandafter \@secondoftwo
 \fi
}%
\providecommand \natexlab [1]{#1}%
\providecommand \enquote  [1]{``#1''}%
\providecommand \bibnamefont  [1]{#1}%
\providecommand \bibfnamefont [1]{#1}%
\providecommand \citenamefont [1]{#1}%
\providecommand \href@noop [0]{\@secondoftwo}%
\providecommand \href [0]{\begingroup \@sanitize@url \@href}%
\providecommand \@href[1]{\@@startlink{#1}\@@href}%
\providecommand \@@href[1]{\endgroup#1\@@endlink}%
\providecommand \@sanitize@url [0]{\catcode `\\12\catcode `\$12\catcode
  `\&12\catcode `\#12\catcode `\^12\catcode `\_12\catcode `\%12\relax}%
\providecommand \@@startlink[1]{}%
\providecommand \@@endlink[0]{}%
\providecommand \url  [0]{\begingroup\@sanitize@url \@url }%
\providecommand \@url [1]{\endgroup\@href {#1}{\urlprefix }}%
\providecommand \urlprefix  [0]{URL }%
\providecommand \Eprint [0]{\href }%
\providecommand \doibase [0]{http://dx.doi.org/}%
\providecommand \selectlanguage [0]{\@gobble}%
\providecommand \bibinfo  [0]{\@secondoftwo}%
\providecommand \bibfield  [0]{\@secondoftwo}%
\providecommand \translation [1]{[#1]}%
\providecommand \BibitemOpen [0]{}%
\providecommand \bibitemStop [0]{}%
\providecommand \bibitemNoStop [0]{.\EOS\space}%
\providecommand \EOS [0]{\spacefactor3000\relax}%
\providecommand \BibitemShut  [1]{\csname bibitem#1\endcsname}%
\let\auto@bib@innerbib\@empty
\bibitem [{\citenamefont {Christensen~et al.}(2019)}]{Christensen2019}%
  \BibitemOpen
  \bibfield  {author} {\bibinfo {author} {\bibfnamefont {J.~E.}\ \bibnamefont
  {Christensen~et al.}},\ }\href@noop {} {\bibfield  {journal} {\bibinfo
  {journal} {arxiv:1907.13331}\ } (\bibinfo {year} {2019})}\BibitemShut
  {NoStop}%
\bibitem [{\citenamefont {Harty~et al.}(2014)}]{Harty2014}%
  \BibitemOpen
  \bibfield  {author} {\bibinfo {author} {\bibfnamefont {T.}~\bibnamefont
  {Harty~et al.}},\ }\href@noop {} {\bibfield  {journal} {\bibinfo  {journal}
  {Phys. Rev. Lett.}\ }\textbf {\bibinfo {volume} {113}},\ \bibinfo {pages}
  {220501} (\bibinfo {year} {2014})}\BibitemShut {NoStop}%
\bibitem [{\citenamefont {Balance~et al.}(2016)}]{Balance2016}%
  \BibitemOpen
  \bibfield  {author} {\bibinfo {author} {\bibfnamefont {C.}~\bibnamefont
  {Balance~et al.}},\ }\href@noop {} {\bibfield  {journal} {\bibinfo  {journal}
  {Phys. Rev. Lett.}\ }\textbf {\bibinfo {volume} {117}},\ \bibinfo {pages}
  {060504} (\bibinfo {year} {2016})}\BibitemShut {NoStop}%
\bibitem [{\citenamefont {Gaebler~et al.}(2016)}]{Gaebler2016}%
  \BibitemOpen
  \bibfield  {author} {\bibinfo {author} {\bibfnamefont {J.}~\bibnamefont
  {Gaebler~et al.}},\ }\href@noop {} {\bibfield  {journal} {\bibinfo  {journal}
  {Phys. Rev. Lett.}\ }\textbf {\bibinfo {volume} {117}},\ \bibinfo {pages}
  {060505} (\bibinfo {year} {2016})}\BibitemShut {NoStop}%
\bibitem [{\citenamefont {Nigg~et al.}(2014)}]{Nigg2014}%
  \BibitemOpen
  \bibfield  {author} {\bibinfo {author} {\bibfnamefont {D.}~\bibnamefont
  {Nigg~et al.}},\ }\href@noop {} {\bibfield  {journal} {\bibinfo  {journal}
  {Science}\ }\textbf {\bibinfo {volume} {345}},\ \bibinfo {pages} {302}
  (\bibinfo {year} {2014})}\BibitemShut {NoStop}%
\bibitem [{\citenamefont {Debnath~et al.}(2016)}]{Debnath2016}%
  \BibitemOpen
  \bibfield  {author} {\bibinfo {author} {\bibfnamefont {S.}~\bibnamefont
  {Debnath~et al.}},\ }\href@noop {} {\bibfield  {journal} {\bibinfo  {journal}
  {Nature}\ }\textbf {\bibinfo {volume} {536}},\ \bibinfo {pages} {63}
  (\bibinfo {year} {2016})}\BibitemShut {NoStop}%
\bibitem [{\citenamefont {Brownnutt~et al.}(2015)}]{Brownnutt2015}%
  \BibitemOpen
  \bibfield  {author} {\bibinfo {author} {\bibfnamefont {M.}~\bibnamefont
  {Brownnutt~et al.}},\ }\href@noop {} {\bibfield  {journal} {\bibinfo
  {journal} {Rev. Mod. Phys.}\ }\textbf {\bibinfo {volume} {87}},\ \bibinfo
  {pages} {1419} (\bibinfo {year} {2015})}\BibitemShut {NoStop}%
\bibitem [{\citenamefont {Talukdar~et al.}(2016)}]{Talukdar2016}%
  \BibitemOpen
  \bibfield  {author} {\bibinfo {author} {\bibfnamefont {I.}~\bibnamefont
  {Talukdar~et al.}},\ }\href@noop {} {\bibfield  {journal} {\bibinfo
  {journal} {Phys. Rev. A}\ }\textbf {\bibinfo {volume} {93}},\ \bibinfo
  {pages} {04315} (\bibinfo {year} {2016})}\BibitemShut {NoStop}%
\bibitem [{\citenamefont {Eltony~et al.}(2016)}]{Eltony2016}%
  \BibitemOpen
  \bibfield  {author} {\bibinfo {author} {\bibfnamefont {A.}~\bibnamefont
  {Eltony~et al.}},\ }\href@noop {} {\bibfield  {journal} {\bibinfo  {journal}
  {Quant. Inf. Proc.}\ }\textbf {\bibinfo {volume} {15}},\ \bibinfo {pages}
  {5351} (\bibinfo {year} {2016})}\BibitemShut {NoStop}%
\bibitem [{\citenamefont {Sedlacek~et al.}(2018)}]{Sedlacek2018}%
  \BibitemOpen
  \bibfield  {author} {\bibinfo {author} {\bibfnamefont {J.}~\bibnamefont
  {Sedlacek~et al.}},\ }\href@noop {} {\bibfield  {journal} {\bibinfo
  {journal} {Phys. Rev. A}\ }\textbf {\bibinfo {volume} {97}},\ \bibinfo
  {pages} {020302(R)} (\bibinfo {year} {2018})}\BibitemShut {NoStop}%
\bibitem [{\citenamefont {Landsman~et al.}(2019)}]{Landsman2019}%
  \BibitemOpen
  \bibfield  {author} {\bibinfo {author} {\bibfnamefont {K.}~\bibnamefont
  {Landsman~et al.}},\ }\href@noop {} {\bibfield  {journal} {\bibinfo
  {journal} {arxiv:1905.10421}\ } (\bibinfo {year} {2019})}\BibitemShut
  {NoStop}%
\bibitem [{\citenamefont {Hudson}\ and\ \citenamefont
  {Campbell}(2018)}]{Hudson2018}%
  \BibitemOpen
  \bibfield  {author} {\bibinfo {author} {\bibfnamefont {E.}~\bibnamefont
  {Hudson}}\ and\ \bibinfo {author} {\bibfnamefont {W.}~\bibnamefont
  {Campbell}},\ }\href@noop {} {\bibfield  {journal} {\bibinfo  {journal}
  {Phys. Rev. A}\ }\textbf {\bibinfo {volume} {98}},\ \bibinfo {pages}
  {040302(R)} (\bibinfo {year} {2018})}\BibitemShut {NoStop}%
\bibitem [{\citenamefont {Stollenwork~et al.}(2018)}]{Stollenwerk2018}%
  \BibitemOpen
  \bibfield  {author} {\bibinfo {author} {\bibfnamefont {P.}~\bibnamefont
  {Stollenwork~et al.}},\ }\href@noop {} {\bibfield  {journal} {\bibinfo
  {journal} {Atoms}\ }\textbf {\bibinfo {volume} {6}},\ \bibinfo {pages} {53}
  (\bibinfo {year} {2018})}\BibitemShut {NoStop}%
\bibitem [{\citenamefont {Schmidt}\ \emph {et~al.}(2005)\citenamefont
  {Schmidt}, \citenamefont {Rosenband}, \citenamefont {Langer}, \citenamefont
  {Itano}, \citenamefont {Bergquist},\ and\ \citenamefont
  {Wineland}}]{Schmidt2005Spectroscopy}%
  \BibitemOpen
  \bibfield  {author} {\bibinfo {author} {\bibfnamefont {P.~O.}\ \bibnamefont
  {Schmidt}}, \bibinfo {author} {\bibfnamefont {T.}~\bibnamefont {Rosenband}},
  \bibinfo {author} {\bibfnamefont {C.}~\bibnamefont {Langer}}, \bibinfo
  {author} {\bibfnamefont {W.~M.}\ \bibnamefont {Itano}}, \bibinfo {author}
  {\bibfnamefont {J.~C.}\ \bibnamefont {Bergquist}}, \ and\ \bibinfo {author}
  {\bibfnamefont {D.~J.}\ \bibnamefont {Wineland}},\ }\href@noop {} {\bibfield
  {journal} {\bibinfo  {journal} {Science}\ }\textbf {\bibinfo {volume}
  {309}},\ \bibinfo {pages} {749} (\bibinfo {year} {2005})}\BibitemShut
  {NoStop}%
\bibitem [{\citenamefont {Wolf~et al.}(2016)}]{Wolf2016}%
  \BibitemOpen
  \bibfield  {author} {\bibinfo {author} {\bibfnamefont {F.}~\bibnamefont
  {Wolf~et al.}},\ }\href@noop {} {\bibfield  {journal} {\bibinfo  {journal}
  {Nature}\ }\textbf {\bibinfo {volume} {530}},\ \bibinfo {pages} {457}
  (\bibinfo {year} {2016})}\BibitemShut {NoStop}%
\bibitem [{\citenamefont {Chou~et al.}(2017)}]{Chou2017}%
  \BibitemOpen
  \bibfield  {author} {\bibinfo {author} {\bibfnamefont {C.-W.}\ \bibnamefont
  {Chou~et al.}},\ }\href@noop {} {\bibfield  {journal} {\bibinfo  {journal}
  {Nature}\ }\textbf {\bibinfo {volume} {545}},\ \bibinfo {pages} {203}
  (\bibinfo {year} {2017})}\BibitemShut {NoStop}%
\bibitem [{\citenamefont {James}(1998)}]{James1998}%
  \BibitemOpen
  \bibfield  {author} {\bibinfo {author} {\bibfnamefont {D.~F.~V.}\
  \bibnamefont {James}},\ }\href@noop {} {\bibfield  {journal} {\bibinfo
  {journal} {Appl. Phys. B}\ }\textbf {\bibinfo {volume} {66}},\ \bibinfo
  {pages} {181} (\bibinfo {year} {1998})}\BibitemShut {NoStop}%
\bibitem [{\citenamefont {Brown}\ \emph {et~al.}(1979)\citenamefont {Brown},
  \citenamefont {Colbourn}, \citenamefont {Watson},\ and\ \citenamefont
  {Wayne}}]{Brown1979Effective}%
  \BibitemOpen
  \bibfield  {author} {\bibinfo {author} {\bibfnamefont {J.~M.}\ \bibnamefont
  {Brown}}, \bibinfo {author} {\bibfnamefont {E.~A.}\ \bibnamefont {Colbourn}},
  \bibinfo {author} {\bibfnamefont {J.~K.~G.}\ \bibnamefont {Watson}}, \ and\
  \bibinfo {author} {\bibfnamefont {F.~D.}\ \bibnamefont {Wayne}},\ }\href@noop
  {} {\bibfield  {journal} {\bibinfo  {journal} {J Molec. Spec.}\ }\textbf
  {\bibinfo {volume} {74}},\ \bibinfo {pages} {294} (\bibinfo {year}
  {1979})}\BibitemShut {NoStop}%
\bibitem [{\citenamefont {Jaynes}\ and\ \citenamefont
  {Cummings}(1963)}]{Jaynes1963}%
  \BibitemOpen
  \bibfield  {author} {\bibinfo {author} {\bibfnamefont {E.}~\bibnamefont
  {Jaynes}}\ and\ \bibinfo {author} {\bibfnamefont {F.}~\bibnamefont
  {Cummings}},\ }\href@noop {} {\bibfield  {journal} {\bibinfo  {journal}
  {Proc. IEEE}\ }\textbf {\bibinfo {volume} {51}},\ \bibinfo {pages} {89}
  (\bibinfo {year} {1963})}\BibitemShut {NoStop}%
\bibitem [{\citenamefont {Schuster~et al.}(2011)}]{Schuster2011}%
  \BibitemOpen
  \bibfield  {author} {\bibinfo {author} {\bibfnamefont {D.}~\bibnamefont
  {Schuster~et al.}},\ }\href@noop {} {\bibfield  {journal} {\bibinfo
  {journal} {Phys. Rev. A}\ }\textbf {\bibinfo {volume} {83}},\ \bibinfo
  {pages} {012311} (\bibinfo {year} {2011})}\BibitemShut {NoStop}%
\bibitem [{\citenamefont {James}\ and\ \citenamefont
  {Jerke}(2007)}]{James2007effective}%
  \BibitemOpen
  \bibfield  {author} {\bibinfo {author} {\bibfnamefont {D.~F.~V.}\
  \bibnamefont {James}}\ and\ \bibinfo {author} {\bibfnamefont
  {J.}~\bibnamefont {Jerke}},\ }\href@noop {} {\bibfield  {journal} {\bibinfo
  {journal} {Can. J. Phys.}\ }\textbf {\bibinfo {volume} {85}},\ \bibinfo
  {pages} {625} (\bibinfo {year} {2007})}\BibitemShut {NoStop}%
\bibitem [{\citenamefont {Senko}\ \emph {et~al.}(2015)\citenamefont {Senko},
  \citenamefont {Richerme}, \citenamefont {Smith}, \citenamefont {Lee},
  \citenamefont {Cohen}, \citenamefont {Retzker},\ and\ \citenamefont
  {Monroe}}]{Senko2015Realization}%
  \BibitemOpen
  \bibfield  {author} {\bibinfo {author} {\bibfnamefont {C.}~\bibnamefont
  {Senko}}, \bibinfo {author} {\bibfnamefont {P.}~\bibnamefont {Richerme}},
  \bibinfo {author} {\bibfnamefont {J.}~\bibnamefont {Smith}}, \bibinfo
  {author} {\bibfnamefont {A.}~\bibnamefont {Lee}}, \bibinfo {author}
  {\bibfnamefont {I.}~\bibnamefont {Cohen}}, \bibinfo {author} {\bibfnamefont
  {A.}~\bibnamefont {Retzker}}, \ and\ \bibinfo {author} {\bibfnamefont
  {C.}~\bibnamefont {Monroe}},\ }\href {\doibase 10.1103/PhysRevX.5.021026}
  {\bibfield  {journal} {\bibinfo  {journal} {Phys. Rev. X}\ }\textbf {\bibinfo
  {volume} {5}},\ \bibinfo {pages} {021026} (\bibinfo {year}
  {2015})}\BibitemShut {NoStop}%
\bibitem [{\citenamefont {Wall}\ \emph {et~al.}(2017)\citenamefont {Wall},
  \citenamefont {Safavi-Naini},\ and\ \citenamefont
  {Rey}}]{Wall2017Bosonmediated}%
  \BibitemOpen
  \bibfield  {author} {\bibinfo {author} {\bibfnamefont {M.~L.}\ \bibnamefont
  {Wall}}, \bibinfo {author} {\bibfnamefont {A.}~\bibnamefont {Safavi-Naini}},
  \ and\ \bibinfo {author} {\bibfnamefont {A.~M.}\ \bibnamefont {Rey}},\ }\href
  {\doibase 10.1103/PhysRevA.95.013602} {\bibfield  {journal} {\bibinfo
  {journal} {Phys. Rev. A}\ }\textbf {\bibinfo {volume} {95}},\ \bibinfo
  {pages} {013602} (\bibinfo {year} {2017})}\BibitemShut {NoStop}%
\bibitem [{\citenamefont {M{\o}lmer}\ and\ \citenamefont
  {S{\o}rensen}(1999)}]{MolmerSorensen99Multiparticle}%
  \BibitemOpen
  \bibfield  {author} {\bibinfo {author} {\bibfnamefont {K.}~\bibnamefont
  {M{\o}lmer}}\ and\ \bibinfo {author} {\bibfnamefont {A.}~\bibnamefont
  {S{\o}rensen}},\ }\href@noop {} {\bibfield  {journal} {\bibinfo  {journal}
  {Phys. Rev. Lett.}\ }\textbf {\bibinfo {volume} {82}},\ \bibinfo {pages}
  {1835} (\bibinfo {year} {1999})}\BibitemShut {NoStop}%
\bibitem [{\citenamefont {Porras}\ and\ \citenamefont
  {Cirac}(2004)}]{Porras2004Effective}%
  \BibitemOpen
  \bibfield  {author} {\bibinfo {author} {\bibfnamefont {D.}~\bibnamefont
  {Porras}}\ and\ \bibinfo {author} {\bibfnamefont {J.~I.}\ \bibnamefont
  {Cirac}},\ }\href@noop {} {\bibfield  {journal} {\bibinfo  {journal} {Phys.
  Rev. Lett.}\ }\textbf {\bibinfo {volume} {92}},\ \bibinfo {pages} {207901}
  (\bibinfo {year} {2004})}\BibitemShut {NoStop}%
\bibitem [{\citenamefont {{D'H}elon}\ and\ \citenamefont
  {Milburn}(1996)}]{DHelon1996Measurements}%
  \BibitemOpen
  \bibfield  {author} {\bibinfo {author} {\bibfnamefont {C.}~\bibnamefont
  {{D'H}elon}}\ and\ \bibinfo {author} {\bibfnamefont {G.~J.}\ \bibnamefont
  {Milburn}},\ }\href@noop {} {\bibfield  {journal} {\bibinfo  {journal} {Phys.
  Rev. A}\ }\textbf {\bibinfo {volume} {54}},\ \bibinfo {pages} {5141}
  (\bibinfo {year} {1996})}\BibitemShut {NoStop}%
\bibitem [{\citenamefont {Cossel~et al.}(2012)}]{Cossel2012}%
  \BibitemOpen
  \bibfield  {author} {\bibinfo {author} {\bibfnamefont {K.}~\bibnamefont
  {Cossel~et al.}},\ }\href@noop {} {\bibfield  {journal} {\bibinfo  {journal}
  {Chem. Phys. Lett.}\ }\textbf {\bibinfo {volume} {546}},\ \bibinfo {pages}
  {1} (\bibinfo {year} {2012})}\BibitemShut {NoStop}%
\bibitem [{\citenamefont {Gresh~et al.}(2016)}]{Gresh2016}%
  \BibitemOpen
  \bibfield  {author} {\bibinfo {author} {\bibfnamefont {D.}~\bibnamefont
  {Gresh~et al.}},\ }\href@noop {} {\bibfield  {journal} {\bibinfo  {journal}
  {J. Mol. Spec.}\ }\textbf {\bibinfo {volume} {319}},\ \bibinfo {pages} {1}
  (\bibinfo {year} {2016})}\BibitemShut {NoStop}%
\bibitem [{\citenamefont {Nielsen}\ and\ \citenamefont
  {Shaffer}(1943)}]{Nielsen1943}%
  \BibitemOpen
  \bibfield  {author} {\bibinfo {author} {\bibfnamefont {H.}~\bibnamefont
  {Nielsen}}\ and\ \bibinfo {author} {\bibfnamefont {W.}~\bibnamefont
  {Shaffer}},\ }\href@noop {} {\bibfield  {journal} {\bibinfo  {journal} {J.
  Chem. Phys.}\ }\textbf {\bibinfo {volume} {11}},\ \bibinfo {pages} {140}
  (\bibinfo {year} {1943})}\BibitemShut {NoStop}%
\bibitem [{\citenamefont {Kozyryev}\ and\ \citenamefont
  {Hutzler}(2017)}]{Kozyryev2017}%
  \BibitemOpen
  \bibfield  {author} {\bibinfo {author} {\bibfnamefont {I.}~\bibnamefont
  {Kozyryev}}\ and\ \bibinfo {author} {\bibfnamefont {N.}~\bibnamefont
  {Hutzler}},\ }\href@noop {} {\bibfield  {journal} {\bibinfo  {journal} {Phys.
  Rev. Lett.}\ }\textbf {\bibinfo {volume} {119}},\ \bibinfo {pages} {133002}
  (\bibinfo {year} {2017})}\BibitemShut {NoStop}%
\bibitem [{\citenamefont {Puri~et al.}(2017)}]{Puri2017}%
  \BibitemOpen
  \bibfield  {author} {\bibinfo {author} {\bibfnamefont {P.}~\bibnamefont
  {Puri~et al.}},\ }\href@noop {} {\bibfield  {journal} {\bibinfo  {journal}
  {Science}\ }\textbf {\bibinfo {volume} {357}},\ \bibinfo {pages} {1370}
  (\bibinfo {year} {2017})}\BibitemShut {NoStop}%
\bibitem [{\citenamefont {O'Rourke}\ and\ \citenamefont
  {Hutzler}(2019)}]{Orourke2019}%
  \BibitemOpen
  \bibfield  {author} {\bibinfo {author} {\bibfnamefont {M.}~\bibnamefont
  {O'Rourke}}\ and\ \bibinfo {author} {\bibfnamefont {N.}~\bibnamefont
  {Hutzler}},\ }\href@noop {} {\bibfield  {journal} {\bibinfo  {journal}
  {arXiv:1902.10683}\ } (\bibinfo {year} {2019})}\BibitemShut {NoStop}%
\bibitem [{\citenamefont {Lien~et al.}(2014)}]{Lien2014}%
  \BibitemOpen
  \bibfield  {author} {\bibinfo {author} {\bibfnamefont {C.-Y.}\ \bibnamefont
  {Lien~et al.}},\ }\href@noop {} {\bibfield  {journal} {\bibinfo  {journal}
  {Nature Comm.}\ }\textbf {\bibinfo {volume} {5}},\ \bibinfo {pages} {4783}
  (\bibinfo {year} {2014})}\BibitemShut {NoStop}%
\end{thebibliography}%

\end{document}